\documentclass[aps,prmaterials,twocolumn,longbibliography,nobibnotes,showkeys,superscriptaddress]{revtex4-2}

\usepackage{graphicx}
\usepackage{dcolumn}
\usepackage{bm}
\usepackage[version=3]{mhchem}

\begin{document}
\title{Why is it so difficult to realize Dy$^{4+}$ in as-synthesized BaZrO$_3$?}
\author{Khang Hoang}
\email[Corresponding author. E-mail: ]{khang.hoang@ndsu.edu}
\affiliation{Center for Computationally Assisted Science and Technology \& Department of Physics, North Dakota State University, Fargo, North Dakota 58108, United States}
\author{Camille Latouche}
\author{St\'{e}phane Jobic}
\affiliation{Institut des Mat\'{e}riaux Jean Rouxel (IMN), Universit\'{e} de Nantes, CNRS, 2 rue de la Houssini\`{e}re, 44322 Nantes, France}

\date{\today}

\begin{abstract}

Rare-earth doped barium zirconate (BaZrO$_3$) ceramics are of interest as proton-conducting and luminescent materials. Here, we report a study of dysprosium (Dy) and other relevant point defects in BaZrO$_3$ using hybrid density-functional defect calculations. The tetravalent Dy$^{4+}$ is found to be structurally and electronically stable at the Zr lattice site (i.e., as Dy$_{\rm Zr}^0$), but most often energetically less favorable than the trivalent Dy$^{3+}$ (i.e., Dy$_{\rm Zr}^-$) in as-synthesized BaZrO$_3$, due to the formation of low-energy, positively charged oxygen vacancies and the mixed-site occupancy of Dy in the host lattice. The Dy$^{4+}$/Dy$^{3+}$ ratio can, in principle, be increased by preparing the material under highly oxidizing and Ba-rich conditions and co-doping with acceptor-like impurities; however, post-synthesis treatment may still be needed to realize a non-negligible Dy$^{4+}$ concentration. We also find that certain unoccupied Dy $4f$ states and the O $2p$ states are {\it strongly hybridized}, a feature not often seen in rare-earth-containing materials, and that the isolated Dy$_{\rm Zr}$ defect might be the source of a broad blue emission in band-to-defect (``charge-transfer'') luminescence.

\end{abstract}

\pacs{}

\maketitle


\section{Introduction}\label{sec;intro}

Dysprosium (Dy) is very stable as the trivalent Dy$^{3+}$ in solid compounds, yet the divalent Dy$^{2+}$ has also been observed in some fluorides \cite{McClure1963JCP,Kiss1965PR} and suspected to be photogenerated under irradiation in persistent luminescent phosphors \cite{Joos2020PRL}. The tetravalent Dy$^{4+}$ was almost unknown up to very recently \cite{Han2012AM,Ricote2018SSI,Ricote2019SSI}. Han et al.~\cite{Han2012AM} was the first to claim experimental evidence of Dy$^{4+}$ in BaZrO$_3$, a perovskite structure type host lattice commonly investigated for its proton conductivity ability \cite{RegaladoVera2021JPE}. Subsequent reports provided further evidence for the presence of the tetravalent state in samples prepared in oxidizing atmospheres \cite{Ricote2018SSI,Ricote2019SSI} and the mixed-site occupancy of Dy in the host lattice \cite{Han2014JACS}, and discussed the possible role of Dy$^{4+}$ as a charge-compensating defect in the oxidation reaction \cite{Ricote2018SSI,Ricote2019SSI}. The stability of Dy$^{4+}$ is thus of interest because of not just the exotic valence state but also the implication it has for defect-related processes in the material.     

On the theory side, the interaction between the Dy dopant and BaZrO$_3$, including native point defects and other impurities that may be present in the host material, has not been investigated, and thus a theoretical foundation for understanding the experimental observations is still lacking. First-principles defect calculations \cite{Freysoldt2014RMP} based on a hybrid density-functional theory (DFT)/Hartree-Fock approach \cite{heyd:8207} can provide a detailed understanding of the atomic and electronic structure (including the oxidation state), energetics, and optical properties of rare-earth (RE) dopants in solid compounds \cite{Hoang2015RRL,Hoang2021PRM}. 

Here, we present an investigation of Dy and other relevant defects in BaZrO$_3$ using hybrid density-functional defect calculations. Explicit calculations are carried out for native point defects and substitutional Dy and yttrium (Y) impurities. Based on the results, we discuss the stability of Dy$^{3+}$ and Dy$^{4+}$ and the tuning of the Dy$^{4+}$/Dy$^{3+}$ ratio and the lattice site preference via tailoring the synthesis conditions and co-doping, or via post-synthesis treatment. Possible band-to-defect and defect-to-band (also known as ``charge-transfer'') optical transitions involving the isolated Dy$_{\rm Zr}$ defect are also explored.

\section{Methodology}\label{sec;method} 

Defects in BaZrO$_3$ are modeled using a supercell approach in which a defect is included in a periodically repeated finite volume of the host material. Here, the term ``defect'' is referred generally to either native point defects or impurities, i.e., extrinsic defects, intentionally added (i.e., dopants) or unintentionally present in the material. A defect is characterized by its formation energy. Defects with a lower formation energy will be more likely to form and occur with a higher concentration. 

The formation energy of a defect X in effective charge state $q$ (with respect to the host lattice) is defined as \cite{Freysoldt2014RMP}     
\begin{align}\label{eq:eform}
E^f({\mathrm{X}}^q)&=&E_{\mathrm{tot}}({\mathrm{X}}^q)-E_{\mathrm{tot}}({\mathrm{bulk}}) -\sum_{i}{n_i\mu_i} \\ %
\nonumber &&+~q(E_{\mathrm{v}}+\mu_{e})+ \Delta^q ,
\end{align}
where $E_{\mathrm{tot}}(\mathrm{X}^{q})$ and $E_{\mathrm{tot}}(\mathrm{bulk})$ are the total energies of the defect-containing and bulk (i.e., perfect and undoped) supercells; $n_{i}$ is the number of atoms of species $i$ that have been added ($n_{i}>0$) or removed ($n_{i}<0$) to form the defect; $\mu_{i}$ is the atomic chemical potential, representing the energy of the reservoir with which atoms are being exchanged. $\mu_{e}$ is the electronic chemical potential, i.e., the Fermi level, representing the energy of the electron reservoir, referenced to the valence-band maximum (VBM) in the bulk ($E_{\mathrm{v}}$). Finally, $\Delta^q$ is the correction term to align the electrostatic potentials of the bulk and defect supercells and to account for finite-size effects on the total energy of charged defects \cite{Freysoldt,Freysoldt11}. 

From the defect formation energy, one can determine the thermodynamic transition level between charge states $q$ and $q'$ of a defect, $\epsilon(q/q')$ ($q>q'$), defined as the Fermi-level position at which the formation energy of the defect in charge state $q$ is equal to that in state $q'$. This level [also referred to as the $(q/q')$ level], corresponding to a {\it defect level}, would be observed in experiments where the defect in the final charge state $q'$ fully relaxes to its equilibrium configuration after the transition. The optical transition level $E_{\rm opt}^{q/q'}$, on the other hand, is defined similarly but with the total energy of the final state $q'$ ($q$) calculated using the lattice configuration of the initial state $q$ ($q'$) for the absorption (emission) process \cite{Freysoldt2014RMP}.

The total-energy electronic structure calculations are based on DFT with the Heyd-Scuseria-Ernzerhof (HSE) functional \cite{heyd:8207}, the projector augmented wave method \cite{PAW1}, and a plane-wave basis set, as implemented in the Vienna {\it Ab Initio} Simulation Package (\textsc{vasp}) \cite{VASP2}. The Hartree-Fock mixing parameter and the screening length are set to the default values of 0.25 and 10 {\AA}, respectively. We use the PAW potentials in the \textsc{vasp} database which treat Ba $5s^{2}5p^{6}6s^{2}$, Zr $4s^{2}4p^{6}5s^{2}4d^{2}$, O $2s^{2}2p^{4}$, Dy $5s^{2}5p^{6}4f^{10}6s^{2}$, and Y $4s^{2}4p^{6}5s^{2}4d^{1}$ explicitly as valence electrons and the rest as core electrons. Defects are simulated using a cubic 40-atom supercell and a 2$\times$2$\times$2 Monkhorst-Pack $k$-point mesh for the integrations over the Brillouin zone. In the defect calculations, the lattice parameters are fixed to the calculated bulk values but all the internal coordinates are relaxed. In all the calculations, the energy cutoff is set to 500 eV and spin polarization is included; structural relaxations are performed with HSE and the force threshold is chosen to be 0.01 eV/{\AA}. The DFT$+$$U$ method \cite{Dudarev1998}, with $U^{\rm eff} = 6.0$ eV applied on the Dy $4f$ states, is also employed to calculate the electronic structure of Dy-doped BaZrO$_3$, and the result is compared with that from the HSE calculations (see Sec.~\ref{sec;results;dy}). For a detailed discussion of the suitability of the hybrid DFT/Hartree-Fock approach to the study of defect physics in RE-doped materials, in comparison with other methods such as DFT and DFT$+$$U$, see Ref.~\citenum{Hoang2021PRM}.

The chemical potentials of Ba, Zr, Dy, Y, and O are referenced to the total energy per atom of bulk Ba, Zr, Dy, Y, and O$_2$ at 0 K, respectively. $\mu_{\rm Ba}$, $\mu_{\rm Zr}$, and $\mu_{\rm O}$ vary over a range determined by the formation enthalpy of BaZrO$_3$ such that $\mu_{\rm Ba} + \mu_{\rm Zr} + 3\mu_{\rm O} = \Delta H({\rm BaZrO_3})$ (calculated to be $-$17.29 eV at 0 K). We examine defect landscape in BaZrO$_3$ in two limits: (i) the (Ba,O)-rich condition where the host material is assumed to be in thermodynamic equilibrium with BaO ($\Delta H = -5.09$ eV) and O$_2$ gas at 1 atm, and (ii) the (Zr,O)-rich condition where equilibrium with ZrO$_2$ ($\Delta H = -10.99$ eV) and O$_2$ gas at 1 atm is assumed. The temperature range from 1000$^\circ$C to 1600$^\circ$C is often used in the preparation of Dy-doped BaZrO$_3$ samples \cite{Han2012AM,Han2014JACS,Ricote2018SSI,Ricote2019SSI}. Under the mentioned conditions, $\mu_{\rm O}$ is $-1.45$ eV (at 1000$^\circ$C) or $-2.56$ eV (1600$^\circ$C) \cite{stull1971}. The specific value of $\mu_{\rm Dy}$ and $\mu_{\rm Y}$ is obtained by assuming equilibrium with Dy$_2$O$_3$ ($\Delta H = -18.53$ eV) and Y$_2$O$_3$ ($\Delta H = -19.00$ eV), respectively. 

In the above, we explicitly take into account the temperature and pressure dependence when considering gas-phase species, specifically when calculating the oxygen chemical potential (which is related to the Gibbs free energy of O$_2$ gas at temperatures and oxygen partial pressures under consideration) \cite{Reuter2001}. Temperature and pressure effects are usually small for solid phases and thus can be ignored; besides, significant cancellation occurs between different terms in the defect formation energy and when comparing different defects \cite{Freysoldt2014RMP}. Also note that the defect transition levels, $\epsilon(q/q')$ and $E_{\rm opt}^{q/q'}$, are {\it independent} of the choice of the atomic chemical potentials.

It should be realized that the sets of the chemical potentials determined from the above assumptions are relevant to the system during synthesis, when {\it all} the constituent elements of the host compound are in exchange with their respective reservoirs. The experimental conditions during, e.g., post-synthesis treatment, measurement, or use are often very different, according to which only certain element(s) of the host may be exchanged with the environment. These conditions, if known, can be translated into other sets of the chemical potentials. For further discussion with examples of defect landscape in different experimental situations in other classes of materials and physical phenomena, see Refs.~\citenum{Hoang2018JPCM} and \citenum{Doux2018ACSAEM}.

\section{Results and discussion}\label{sec;results}

\subsection{Bulk properties and native defects}

In bulk (cubic) BaZrO$_3$, the calculated lattice constant is 4.200 {\AA}, compared to 4.192 {\AA} in experiments \cite{Yamanaka2005JNM}. The calculated band gap is 4.68 eV, an indirect gap with the VBM at the $R$ ($\frac{1}{2},\frac{1}{2},\frac{1}{2}$) point in the reciprocal space and the conduction-band minimum (CBM) at the $\Gamma$ (0,0,0) point. For comparison, the reported experimental band gap varies in the range from 4.0 eV to 5.3 eV \cite{Borja-Urby2010MSEB,Yuan2008IJHE,Robertson2000JVST}. The electronic contribution to the static dielectric constant is found to be 3.98 in our HSE calculations, based on the real part of the dielectric function $\epsilon_{1}(\omega)$ for $\omega\rightarrow0$. The ionic contribution is calculated using density-functional perturbation theory \cite{dielectricmethod}, within the generalized-gradient approximation \cite{GGA}. The total dielectric constant is 51.75, in good agreement with the reported experimental value of $\sim$47 at 0 K (or $\sim$52 if taking into account the porosity of the sample \cite{Akbarzadeh2005PRB}). In a simple ionic model, BaZrO$_3$ can be regarded as consisting of Ba$^{2+}$, Zr$^{4+}$, and O$^{2-}$.

\begin{figure}
\vspace{0.2cm}
\includegraphics*[width=\linewidth]{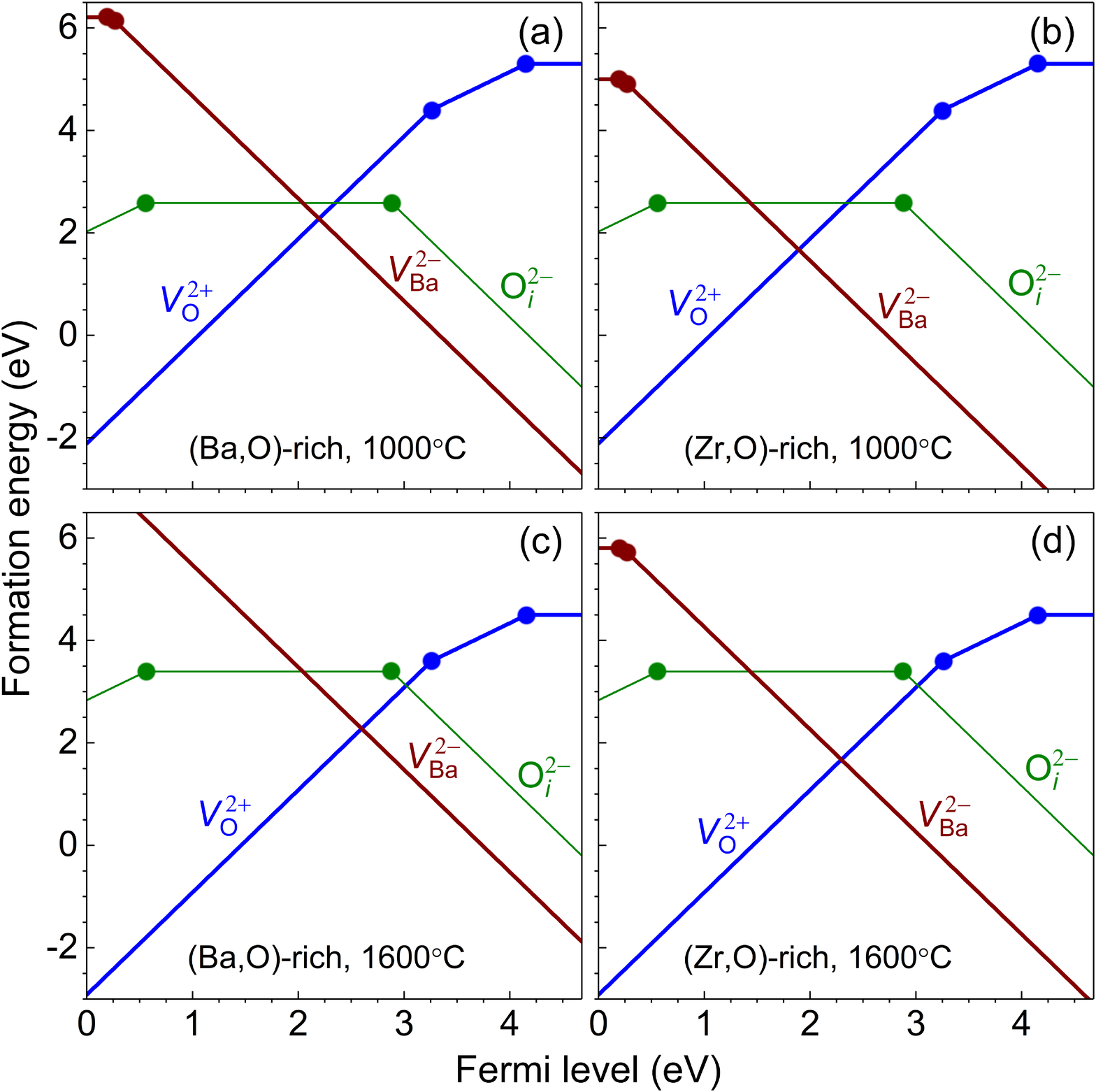}
\caption{Formation energies of selected native defects in BaZrO$_3$ as a function of the Fermi level from the VBM (at 0 eV) to the CBM (4.68 eV), under the (Ba,O)-rich or (Zr,O)-rich condition at 1000$^\circ$C or 1600$^\circ$C. Only segments of the energy line corresponding to the lowest-energy charge states are shown. The slope of these segments indicates the charge state ($q$): positively (negatively) charged defect configurations have positive (negative) slopes; horizontal segments correspond to neutral conﬁgurations. Large solid dots connecting two segments with different slopes mark the defect levels.} 
\label{fig;fe;natives} 
\end{figure}

Figure \ref{fig;fe;natives} shows the formation energy of selected native point defects in BaZrO$_3$. The dominant defects are positively charged oxygen vacancy, $V_{\rm O}^{2+}$ (or, equivalently, the removal of an O$^{2-}$ ion from the host material), and negatively charged barium vacancy, $V_{\rm Ba}^{2-}$ (the removal of a Ba$^{2+}$ from the host). In the absence of impurities, intentionally incorporated or unintentionally present, the Fermi level is ``pinned'' at the position where $V_{\rm O}^{2+}$ and $V_{\rm Ba}^{2-}$ have equal formation energies and hence charge neutrality is maintained. Other defects, such as the oxygen interstitial (O$_i$), have a higher formation energy. In O$_i$, the extra O combines with one of the O atoms of the host to form an O--O dumbbell that is perpendicular to the line connecting the two Ba atoms in the original Ba--O--Ba chain. Our results for the native defects are in agreement with those reported by Rowberg et al.~\cite{Rowberg2019ACSAEM}, except that we find a lower-energy configuration for O$_i^0$.      

\subsection{Doping with dysprosium} \label{sec;results;dy}

\begin{figure}
\vspace{0.2cm}
\includegraphics*[width=\linewidth]{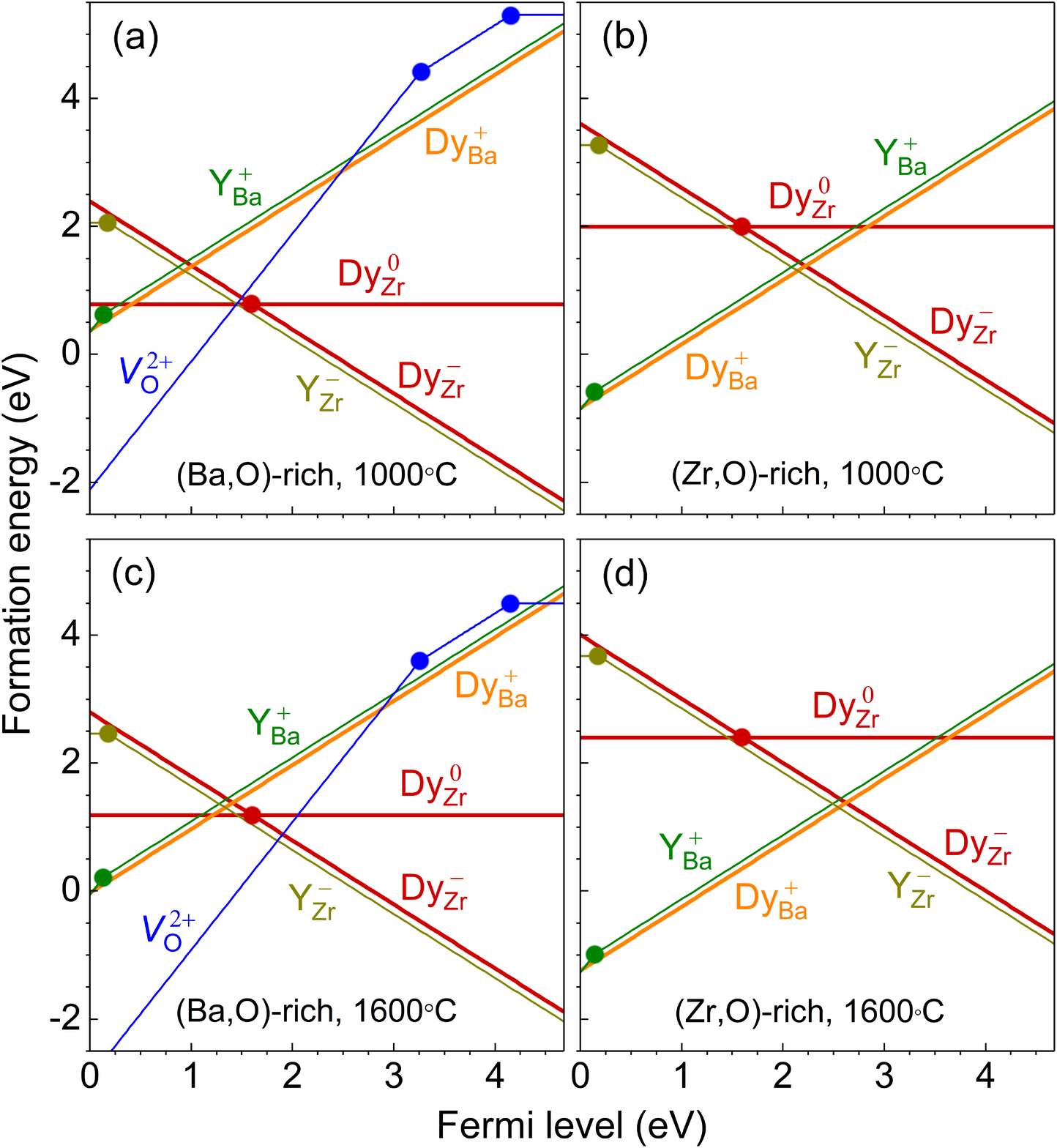}
\caption{Formation energies of Dy- and Y-related defects as a function of the Fermi level from the VBM to the CBM, under the (Ba,O)-rich or (Zr,O)-rich conditions. Large solid dots mark the defect levels. The results for $V_{\rm O}$ under the (Ba,O)-rich condition are repeated here for easy comparison.} 
\label{fig;fe;re} 
\end{figure}

Figure \ref{fig;fe;re} shows the formation energy of the substitutional Dy impurity either at the Ba site (Dy$_{\rm Ba}$) or the Zr site (Dy$_{\rm Zr}$). We find that Dy$_{\rm Ba}$ is stable only as Dy$_{\rm Ba}^{+}$ (i.e., the trivalent Dy$^{3+}$ at the Ba site, with spin $S=5/2$), irrespective of the chosen set of chemical potentials (which represents synthesis conditions). In the  Dy$_{\rm Ba}^{+}$ configuration, the Dy$^{3+}$ ion moves off-center by 0.69 {\AA} and comes closer to six of its twelve neighboring O atoms, with the Dy--O bond length being 2.277--2.512 {\AA}. Dy$_{\rm Zr}$ is, on the other hand, structurally and electronically stable as Dy$_{\rm Zr}^-$ (i.e., the trivalent Dy$^{3+}$ at the Zr site, $S=5/2$) and Dy$_{\rm Zr}^0$ (i.e., the tetravalent Dy$^{4+}$ at the Zr site, $S=3$). The thermodynamic transition level $(0/-)$ of Dy$_{\rm Zr}$ is at 1.61 eV above the VBM; i.e., above (below) this level, Dy$^{3+}$ (Dy$^{4+}$) is energetically more stable (see further discussion later). In the Dy$_{\rm Zr}^0$ (Dy$_{\rm Zr}^-$) configuration, the Dy--O bond is 2.131--2.134 {\AA} (2.188 {\AA}), consistent with the fact that the ionic radius of Dy$^{4+}$ (0.78 {\AA}) \cite{Knop1974CJC} is smaller than that of Dy$^{3+}$ (0.912 {\AA}) \cite{Shannon1976}.

\begin{figure}
\centering
\includegraphics[width=\linewidth]{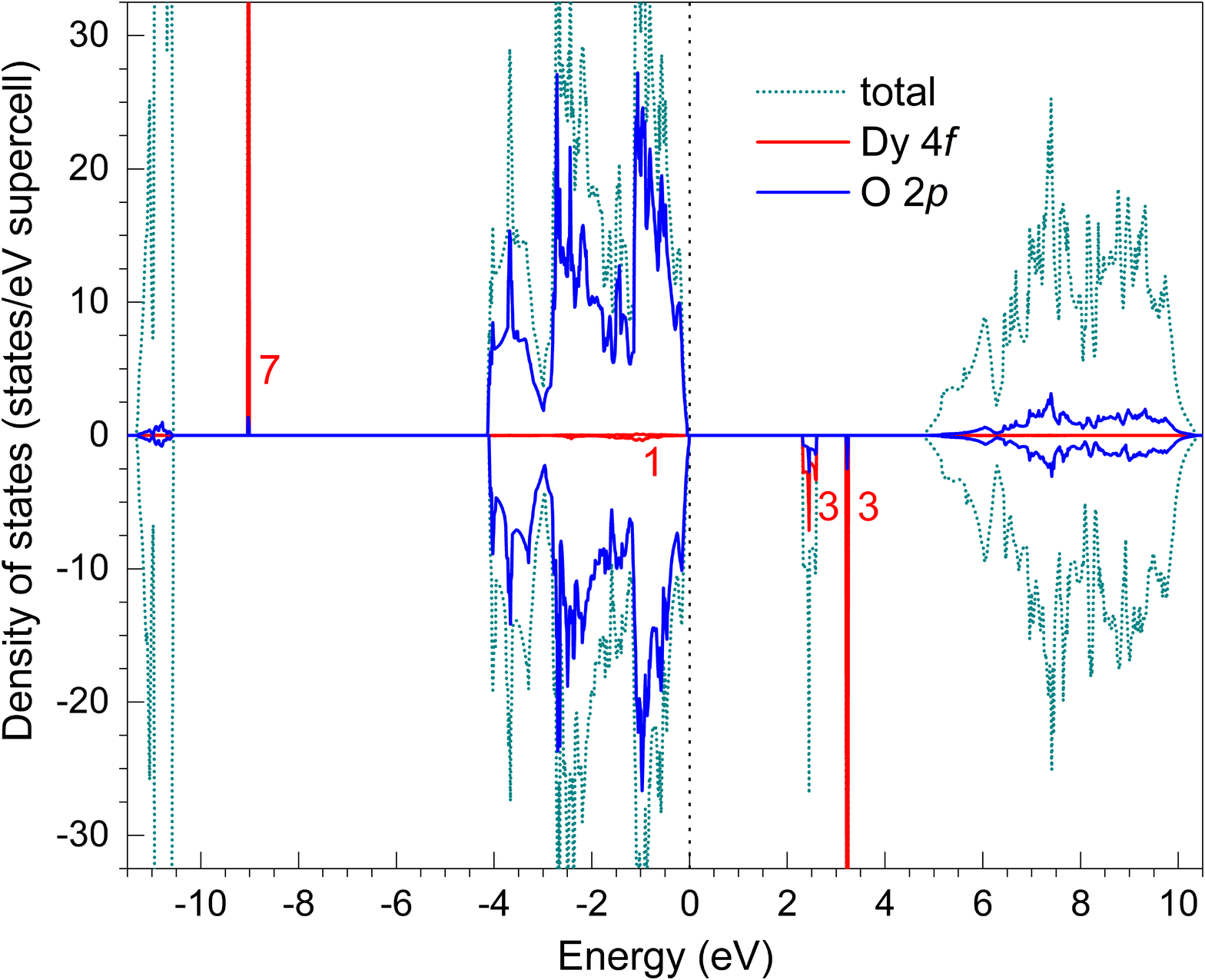}
\caption{Total and projected densities of states (DOS) of Dy-doped BaZrO$_3$, specifically Dy$_{\rm Zr}^0$, obtained in HSE calculations. The spin-majority spectrum is on the $+y$ axis, and the spin-minority spectrum is on the $-y$ axis. The number of $4f$ electrons at the Dy $4f$-projected DOS peaks is indicated. The zero of energy is set to the highest occupied state.}
\label{fig;dos}
\end{figure}

To understand the electronic stability of Dy$^{3+}$ and Dy$^{4+}$, we show in Fig.~\ref{fig;dos} the total and Dy $4f$-projected electronic density of states (DOS) of Dy-doped BaZrO$_3$. In the DOS calculations, one Zr atom in the 40-atom BaZrO$_3$ supercell is substituted with Dy; i.e., the chemical composition is Ba$_{8}$DyZr$_{7}$O$_{24}$, which corresponds to the Dy$_{\rm Zr}^0$ defect configuration discussed earlier. High-quality DOS is achieved by using a $\Gamma$-centered 4$\times$4$\times$4 $k$-point mesh. The ground state of Dy$^{4+}$ is $4f^{8}$. In the calculated DOS, we find that Dy introduces seven spin-up occupied $4f$ states at about $-$9 eV, one spin-down occupied $4f$ state at about $-$0.7 eV (not clearly seen in Fig.~\ref{fig;dos} due to the energy resolution of the calculation but is confirmed by electron counting and by an examination of the wave functions), and six spin-down unoccupied $4f$ states in the host band gap (at about $+$2.5 eV and $+$3.2 eV). In going from Dy$_{\rm Zr}^0$ to Dy$_{\rm Zr}^-$, the lowest unoccupied Dy $4f$ state captures an electron and Dy$^{4+}$ becomes Dy$^{3+}$ ($4f^{9}$). Note that our DFT$+$$U$ calculations produce an electronic structure that is qualitatively similar to that reported in Fig.~\ref{fig;dos}. However, since the Hubbard $U$ term is only applied on the Dy $4f$ states and all other orbitals in the compound are left uncorrected, the host DFT$+$$U$ band gap is significantly underestimated. As a result, the position of the Dy $4f$ states with respect to the band edges are different from that obtained with HSE.

\begin{figure}
\vspace{0.2cm}
\includegraphics*[width=0.65\linewidth]{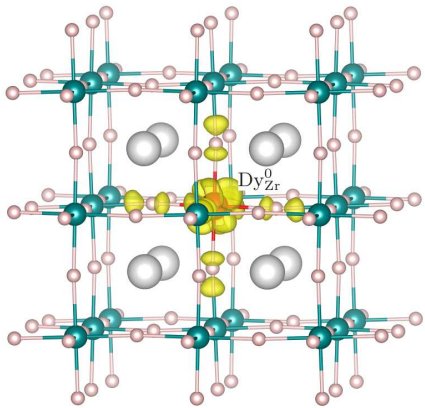}
\caption{Structure of the Dy$_{\rm Zr}^0$ (i.e., Dy$^{4+}$) defect configuration in BaZrO$_3$. The charge density shows an electron hole localized at the Dy ion and the neighboring O sites. The isovalue for the charge-density isosurface (yellow) is set to 0.02 $e$/{\AA}$^3$. The large (gray) spheres are Ba, medium (red/blue) spheres are Dy/Zr, and small (red) spheres are O.} 
\label{fig;fe;struct} 
\end{figure}

Although the RE $4f$ orbitals are often regarded as being well shielded by the outer $5s^2$ and $5p^6$ electron shells and thus contribute very weakly to chemical bonding, we find a {\it strong hybridization} between certain unoccupied Dy $4f$ orbitals and the O $2p$ states. This can be observed in the relatively broad peak at about $+$2.5 eV in the calculated DOS shown in Fig.~\ref{fig;dos} where there are contributions from the O $2p$ states and the structure of the O $2p$-derived DOS peak resembles that of the Dy $4f$-derived one. Figure \ref{fig;fe;struct} shows the lattice geometry of Dy$_{\rm Zr}^0$ and the charge density of the associated electron hole (with respect to Dy$_{\rm Zr}^-$). The charge density, again, shows a strong Dy $4f$--O $2p$ hybridization. Note that the mixing of the RE $4f$ states and the O states has also been reported in the case of Gd$_3$Ga$_5$O$_{12}$ \cite{Juhin2019PRM}.   

It should be noted that the Dy-derived peaks in the DOS (Fig.~\ref{fig;dos}) are not defect levels associated with Dy$_{\rm Zr}$. Indeed, those Kohn-Sham levels cannot directly be identified with any levels that can be observed in experiments \cite{Freysoldt2014RMP}. The $(0/-)$ level of Dy$_{\rm Zr}$, for example, must be calculated using the total energies of the Dy$_{\rm Zr}^0$ and Dy$_{\rm Zr}^-$ configurations as described earlier and reported in Fig.~\ref{fig;fe;re}.

Considering only the Dy-related defects, we find that, under the (Zr,O)-rich condition, Dy$^{3+}$ (in the form of Dy$_{\rm Ba}^+$ and/or Dy$_{\rm Zr}^-$) is energetically more favorable than Dy$^{4+}$ (in the form of Dy$_{\rm Zr}^0$) in the entire range of Fermi-level values; see Figs.~\ref{fig;fe;re}(b) and \ref{fig;fe;re}(d). Under the (Ba,O)-rich condition, Dy$^{4+}$ is energetically more favorable than Dy$^{3+}$ in a small range of Fermi-level values, and this range is greater at lower temperatures; see Figs.~\ref{fig;fe;re}(a) and \ref{fig;fe;re}(c). This is because the lower temperature leads to higher formation energies of $V_{\rm O}^{2+}$ and Dy$_{\rm Ba}^+$ and lower formation energies of Dy$_{\rm Zr}^0$ (as well as Dy$_{\rm Zr}^-$), thus increasing the range of Fermi-level values in which Dy$_{\rm Zr}^0$ is energetically more favorable; the formation energy of Dy$_{\rm Zr}^0$ is 0.78 eV at 1000$^\circ$C, compared to 1.19 eV at 1600$^\circ$C.

That Dy$^{4+}$-favorable range, however, may not even be accessible if one takes into consideration the native defects. Under the (Ba,O)-rich condition at 1600$^\circ$C, for example, the Fermi level is at 1.90 eV, determined predominantly by $V_{\rm O}^{2+}$ and Dy$_{\rm Zr}^-$, at which the formation energy of Dy$_{\rm Zr}^0$ is higher than that of Dy$_{\rm Zr}^-$ by 0.30 eV; see Fig.~\ref{fig;fe;re}(c). Only at lower temperatures, e.g., 1000$^\circ$C, Dy$_{\rm Zr}^0$ has a lower energy than Dy$_{\rm Zr}^-$ (by 0.11 eV, where the Fermi level is at 1.50 eV); see Fig.~\ref{fig;fe;re}(a). Note, however, that an energy difference of $\sim$0.1 eV is within the error bar of our calculations. Yet what is clear here is that low-energy, positively charged defects, e.g., $V_{\rm O}^{2+}$ and Dy$_{\rm Ba}^+$, reduce the Dy$^{4+}$-favorable range or make it inaccessible. The effect of Y co-doping will be discussed later.

Overall, our results indicate that higher Dy$^{4+}$ concentrations might be achieved with higher oxygen ($\mu_{\rm O}$) and barium ($\mu_{\rm Ba}$) chemical potential values [and hence lower zirconium ($\mu_{\rm Zr}$) chemical potential values]. Higher $\mu_{\rm O}$ values represent more oxidizing environments, which are usually associated with oxygen gas at lower temperatures and higher oxygen partial pressures \cite{Reuter2001}, and lead to a reduced oxygen vacancy concentration (particularly, that of $V_{\rm O}^{2+}$) due to an increase in the defect's formation energy. Similarly, higher $\mu_{\rm Ba}$ values, which can be achieved by preparing the material under Ba-rich conditions, result in a reduction or suppression of the Ba-site occupancy (the formation of Dy$_{\rm Ba}^{+}$). Lower $\mu_{\rm Zr}$ values result in a lower formation energy (and hence a higher concentration) of Dy at the Zr site. Note that our argumentation assumes that the material is prepared under (or close to) thermodynamic equilibrium conditions. It is, however, not clear at this point if the defect landscape reported in Fig.~\ref{fig;fe;re}(a) or others that are close to it can be realized under actual synthesis conditions such that the Dy$^{4+}$ concentration is significant, or if post-synthesis treatment is needed to realize Dy$^{4+}$ (see more below).  

The above results showing Dy$_{\rm Ba}^+$ can have a low formation energy are consistent with experimental reports of the mixed-site occupancy in which Dy is incorporated at both the Ba and Zr sites \cite{Han2014JACS}. The synthesis conditions in Han et al.~\cite{Han2014JACS} should be somewhere between the (Ba,O)-rich and (Zr,O)-rich limits discussed in this work.   

\subsection{(Co-)doping with yttrium}

To understand the effects of doping or co-doping with Y, calculations are also carried out for the substitutional Y impurity. Figure \ref{fig;fe;re} shows that Y$_{\rm Ba}$ is most stable as Y$_{\rm Ba}^+$ (i.e., Y$^{3+}$ at the Ba site, $S=0$), except in a very small range of Fermi-level values near the VBM (0 to 0.15 eV) where it is stable as Y$_{\rm Ba}^{2+}$ that is, in fact, a complex of Y$_{\rm Ba}^+$ and a localized hole at an O site (i.e., O$^-$); the Y$^{3+}$ ion in the Y$_{\rm Ba}$ defects is significantly off-center (by 0.73 {\AA} in the case of Y$_{\rm Ba}^+$), similar to Yb$_{\rm Ba}$. Y$_{\rm Zr}$ introduces a defect level, $(0/-)$, at 0.18 eV above the VBM; and the extra electron in Y$_{\rm Ba}^0$ is localized over several neighboring O sites. Like Dy, Y can thus be incorporated at the Zr site and/or the Ba site, depending on the Fermi-level position, which is consistent with the mixed-site occupancy observed in experiments \cite{Azad2008JMC,Yamazaki2010JMC,Han2013JMCA}. Our results for Y are similar those reported by Rowberg et al.~\cite{Rowberg2019ACSAEM}.

When Y is incorporated at the Zr lattice site (thus forming the Y$_{\rm Zr}^-$ defect) with a concentration higher (i.e., a formation energy lower) than that of Dy$_{\rm Zr}^-$, as it is the case shown in Fig.~\ref{fig;fe;re}, the Fermi level will be shifted toward the VBM as the system reestablishes charge neutrality. Under the (Ba,O)-rich condition, for example, the Fermi level is now determined predominantly by $V_{\rm O}^{2+}$ and Y$_{\rm Zr}^-$; see Figs.~\ref{fig;fe;re}(a) and \ref{fig;fe;re}(c). As the Fermi level is shifted leftward, the formation energy of Dy$_{\rm Zr}^-$ increases (and hence its concentration decreases) whereas that of Dy$_{\rm Zr}^0$ remains constant, thus increasing the Dy$^{4+}$/Dy$^{3+}$ ratio. Any acceptor-like defect that shifts the Fermi level would lead to a similar effect. For instance, Rb$_{\rm Ba}$ and Sc$_{\rm Zr}$, stable as Rb$_{\rm Ba}^-$ and Sc$_{\rm Zr}^-$, respectively, as reported in Ref.~\citenum{Rowberg2019ACSAEM}, would be effective if incorporated with a concentration higher than that of Dy$_{\rm Zr}^-$; for an illustration of the Fermi-level shifting effect, see, e.g., Fig.~11(a) of Ref.~\citenum{Hoang2018JPCM}. Note that, even with co-doping, one still need to suppress the formation of oxygen vacancies and the Ba-site occupancy (see above), otherwise $V_{\rm O}^{2+}$ and/or Dy$_{\rm Ba}^{+}$ will counteract the effect of co-doping. Also note that, a Fermi-level shift alone (i.e., without decreasing the formation energy of Dy$_{\rm Zr}^0$) does not increase the Dy$^{4+}$ concentration as Dy$_{\rm Zr}^0$ is a neutral defect and its formation energy is independent of the Fermi-level position; i.e., it only changes the Dy$^{3+}$ concentration and thus the Dy$^{4+}$/Dy$^{3+}$ ratio.

\subsection{Post-synthesis treatment} 

Given the above results and discussion, one may expect that Dy exists entirely or predominantly as the trivalent Dy$^{3+}$ (i.e., Dy$_{\rm Zr}^-$) in BaZrO$_3$ under normal synthesis conditions, and Dy$_{\rm Zr}^-$ is charge-compensated predominantly by $V_{\rm O}^{2+}$ and/or Dy$_{\rm Ba}^+$. The native defects and RE-related defects, once formed or incorporated during the synthesis reaction, are expected to remain trapped in the material after the synthesis and act as athermal, prexisting defects in subsequent experiments \cite{Hoang2018JPCM}. We refer to the BaZrO$_3$ material at this stage as being {\it as-synthesized}.

Upon heat treating as-synthesized BaZrO$_3$ in oxidizing atmospheres, the following reaction is expected:
\begin{equation}\label{eq;ox}
\frac{1}{2}{\rm O}_2 + V_{\rm O}^{2+} + 2{\rm Dy}_{\rm Zr}^- \rightarrow {\rm O}_{\rm O}^0 + 2{\rm Dy}_{\rm Zr}^0,
\end{equation}
in which oxygen from the environment fills the oxygen vacancies and, to maintain charge neutrality, Dy$^{3+}$ (i.e., Dy$_{\rm Zr}^-$) is oxidized to Dy$^{4+}$ (i.e., Dy$_{\rm Zr}^0$). Here, we assume that the system exchanges only oxygen with the environment. It should be emphasized here that reaction (\ref{eq;ox}) is possible because Dy$^{4+}$ (and Dy$^{3+}$) can be stabilized in BaZrO$_3$, even when Dy$^{4+}$ is not already present in as-synthesized samples. Also note that, in principle, the right-hand side of Eq.~(\ref{eq;ox}) can include Dy$_{\rm Zr}^-$ and an electron hole localized at an O site (``O$^-$'') or a free hole ($h^+$) which acts as a charge-compensating defect; however, we find that neither O$^-$ nor $h^+$ can be stabilized when an electron hole is added to the supercell containing Dy$_{\rm Zr}^-$.

The above oxidation reaction may explain the presence of the tetravalent Dy$^{4+}$ reported in BaZrO$_3$-based materials prepared in highly oxidizing atmospheres \cite{Han2012AM,Ricote2018SSI,Ricote2019SSI}. It would be useful if the Dy$^{4+}$/Dy$^{3+}$ ratio and the Fermi-level position of those samples can be determined.  

As reported by Han et al.~\cite{Han2012AM}, Dy$^{4+}$ can be converted into Dy$^{3+}$ again in subsequent heat treatment in highly reducing atmospheres (e.g., H$_2$). Using our defect notation, the reaction can be written as
\begin{equation}\label{eq;red}
{\rm O}_{\rm O}^0 + {\rm H}_2 + 2{\rm Dy}_{\rm Zr}^0 \rightarrow {\rm H}_2{\rm O} + V_{\rm O}^{2+} + 2{\rm Dy}_{\rm Zr}^-.
\end{equation}
Again, this reaction, like reaction (\ref{eq;ox}), is possible because both Dy$^{3+}$ and Dy$^{4+}$ can be stabilized in BaZrO$_3$.

\subsection{Band-to-defect luminescence}

\begin{figure}
\vspace{0.2cm}
\includegraphics*[width=0.9\linewidth]{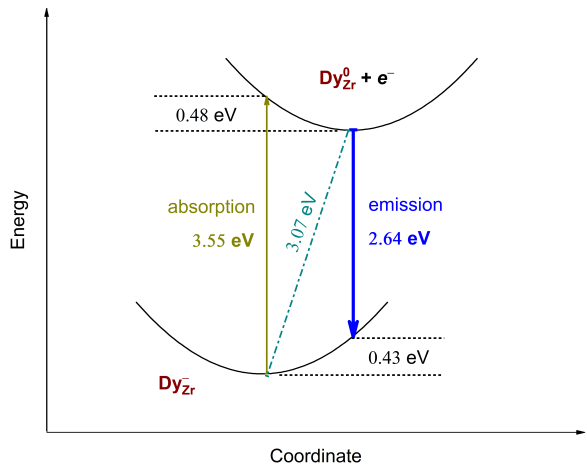}
\caption{Configuration-coordinate diagram illustrating optical emission (down arrow) and absorption (up arrow) processes involving Dy$_{\rm Zr}$ in BaZrO$_3$. The thermal energy [also called the zero-phonon line (ZPL), the dash-dotted line] is the thermodynamic transition level $\epsilon(0/-)$ relative to CBM. The values sandwiched between two dotted lines are the relaxation energies (the Franck-Condon shifts). Axes are not to scale. For a detailed discussion of the configuration-coordinate diagram representation, see, e.g., Alkauskas et al.~\cite{Alkauskas2016JAP}.}
\label{fig;cc} 
\end{figure}

Finally, we examine optical transitions involving the isolated Dy$_{\rm Zr}$ defect in BaZrO$_3$. Given the presence of the $(0/-)$ level in the host band gap, Dy$_{\rm Zr}$ may act as a carrier trap for band-to-defect and defect-to-band optical transitions. Figure \ref{fig;cc} illustrates Dy$_{\rm Zr}$-related optical absorption and emission processes. Under illumination, for example, the Dy$_{\rm Zr}^-$ defect configuration can absorb a photon and become Dy$_{\rm Zr}^0$ with the removed electron being excited into the conduction band. The peak absorption energy related to the optical transition level $E_{\rm opt}^{-/0}$ (the formation energy difference between Dy$_{\rm Zr}^-$ and the Dy$_{\rm Zr}^0$ configuration in the lattice geometry of Dy$_{\rm Zr}^-$) is 3.55 eV, with a relaxation energy (i.e., the Franck-Condo shift) of 0.48 eV. Dy$_{\rm Zr}^0$ can then capture an electron from the CBM, e.g., previously excited from Dy$_{\rm Zr}^-$ (or the valence band) to the conduction band, and emits a photon. The peak emission energy related to the optical transition level $E_{\rm opt}^{0/-}$ (the formation energy difference between Dy$_{\rm Zr}^0$ and the Dy$_{\rm Zr}^-$ configuration in the lattice geometry of Dy$_{\rm Zr}^0$) is 2.64 eV, with a relaxation energy of 0.43 eV. This indicates that Dy$_{\rm Zr}$ can potentially be the source of a broad blue emission associated with the Dy$^{4+}$ $+$ Zr$^{3+}$ $\rightarrow$ Dy$^{3+}$ $+$ Zr$^{4+}$ transition. Further experimental studies are needed to characterize these possible ``charge-transfer'' transitions (as well as Dy $4f$--$4f$ transitions, not considered in this work) in Dy-doped BaZrO$_3$.

\section{Conclusions} 

We have carried out a study of Dy and other relevant defects in BaZrO$_3$ using hybrid density-functional calculations. We find that Dy has mixed-site occupancy and is stable predominantly as the trivalent Dy$^{3+}$ in the bulk material. The tetravalent Dy$^{4+}$ is found to be structurally and electronically stable at the Zr lattice site, but most often energetically less favorable than the trivalent Dy$^{3+}$. This is due to the presence of low-energy, positively charged oxygen vacancies and the mixed-site occupancy of Dy in as-synthesized BaZrO$_3$. The Dy$^{4+}$/Dy$^{3+}$ ratio can, in principle, be increased by preparing the material under highly oxidizing and Ba-rich conditions and co-doping with acceptor-like impurities; however, post-synthesis treatment in oxidizing atmospheres may be needed to realize a non-negligible Dy$^{4+}$ concentration. The lattice site preference of Dy (and Y) can also be tuned by tuning the synthesis conditions. Interestingly, we find a strong hybridization between the Dy $4f$ states and the O $2p$ states and that the isolated substitutional Dy$_{\rm Zr}$ defect can be the source of a broad blue emission in band-to-defect luminescence. Our work thus provides a theoretical foundation for understanding experimental observations as well as guidelines for further studies. It also serves as a methodological template for investigating impurities with suspected exotic valence states.



\begin{acknowledgments}

K.H.~is grateful to Universit\'{e} de Nantes for supporting his visit to Institut des Mat\'{e}riaux Jean Rouxel (IMN) during which this work was initiated. This work used resources of the Center for Computationally Assisted Science and Technology (CCAST) at North Dakota State University, which were made possible in part by NSF MRI Award No.~2019077, and of Centre de Calcul Intensif des Pays de Loire (CCIPL), France. 

\end{acknowledgments}


%

\end{document}